\documentclass[aps,preprint,floats,nofootinbib]{revtex4}
\usepackage{amsmath}
\usepackage{amssymb}
\usepackage{latexsym}
\usepackage{multirow}
\usepackage{longtable}
\usepackage{graphicx}

\newlength{\defbaselineskip}
\newcommand{\setlinespacing}[1]%
           {\setlength{\baselineskip}{#1 \newcommand}}

\def\lsim{\mathrel{\raise.3ex\hbox{$<$\kern-.75em\lower1ex\hbox{$\sim$}}}} 
\def\gsim{\mathrel{\raise.3ex\hbox{$>$\kern-.75em\lower1ex\hbox{$\sim$}}}} 
\begin{document}

\preprint{
\hfill
\begin{minipage}[t]{3in}
\begin{flushright}
\vspace{0.0in}
FERMILAB--PUB--10--387--A\\
\end{flushright}
\end{minipage}
}

\hfill$\vcenter{\hbox{}}$

\vskip 0.5cm

\title{Probing Exotic Physics With Supernova Neutrinos}
\author{Chris Kelso$^1$ and Dan Hooper$^{2,3}$}
\affiliation{$^1$Department of Physics, University of Chicago \\
$^2$Department of Astronomy and Astrophysics, University of Chicago \\
$^3$Theoretical Astrophysics Group, Fermi National Accelerator Laboratory}
\date{\today}

\bigskip

\begin{abstract}
Future galactic supernovae will provide an extremely long baseline for studying the properties and interactions of neutrinos.  In this paper, we discuss the possibility of using such an event to constrain (or discover) the effects of exotic physics in scenarios that are not currently constrained and are not accessible with reactor or solar neutrino experiments. In particular, we focus on the cases of neutrino decay and quantum decoherence. We calculate the expected signal from a core-collapse supernova in both current and future water Cerenkov, scintillating, and liquid argon detectors, and find that such observations will be capable of distinguishing between many of these scenarios. Additionally, future detectors will be capable of making strong, model-independent conclusions by examining events associated with a galactic supernova's neutronization burst. 
\end{abstract}

\maketitle

\newpage

\section{Introduction}

In many scenarios of physics beyond the standard model, the associated signatures are observable only in high energy interactions, motivating the use of particle accelerators such as the Large Hadron Collider and the Tevatron. In other cases, however, the effects of new physics can appear over very long periods of time, or in very long baseline experiments. The discovery of neutrino flavor oscillations provide such an example. Neutrino oscillations have only been observed among neutrinos propagating between the Sun and Earth and over shorter distances, however, leaving open the possibility that other new phenomena might become observable through the study of neutrinos traveling over much greater distances~\cite{longbaseline,longbaseline2}. 

A wide variety of distant astrophysical objects are predicted to potentially produce observable fluxes of neutrinos, including active galactic nuclei, gamma ray bursts, microquasars, X-ray binaries, and supernovae~\cite{neutrinoastro}. Among these prospective sources, however, only neutrinos from supernovae have been observed. In particular, a total of 24 (likely anti-electron) neutrinos were detected by three different detectors in the seconds following supernova 1987A. At a distance of approximately 50 kiloparsecs from Earth, SN 1987A provided an opportunity to observe neutrinos over a baseline that is roughly $10^{10}$ times longer than that traveled by solar neutrinos. When the next nearby supernova takes place, modern detectors will observe far more events and collect a much greater range of information than was possible in 1987. 

In this article, we discuss the possibility of using neutrinos from a galactic supernova as a probe of exotic physics. In particular, we focus on two examples of phenomena that are observable only over long baselines or timescale: neutrino decay and quantum decoherence. The remainder of this article is structured as follows. In Sec.~\ref{physics}, we summarize the exotic physics scenarios considered in our study. In Sec.~\ref{sec:initialFlux}, we discuss the predicted neutrino spectrum from supernova and in Sec.~\ref{detect} we describe the experiments used to detect them. In Sec.~\ref{results}, we describe how neutrino decay or quantum decoherence would impact the neutrinos observed from a galactic supernova and discuss how such observations could be used to constrain or identify such phenomena. In Sec.~\ref{neutronization}, we discuss the role that neutrinos from a supernova's neutronization burst can play in identifying or constraining new physics. Finally, in Sec.~\ref{conclusions}, we summarize our results and conclusions.

\section{Exotic Physic Scenarios}
\label{physics}

\subsection{Neutrino Decay}

\noindent
In this study, we focus on two body decays of the form
\begin{equation}
\nu_i\rightarrow\nu_j+X,
\label{eq:decay}
\end{equation} 
where $\nu_i$ and $\nu_j$ denote either neutrino or anti-neutrino mass eigenstates and $X$ represents a very light or massless particle (such as a majoran) that is not detected (for related model building, see Ref.~\cite{decaymodels}). In contrast to radiative decays~\cite{radiative}, decays of this form are viable over supernova-scale baselines. The strongest constraints on decays of this type are on the order of $\tau/m\gsim10^{-4}\,$s/eV, and result from solar neutrino data~\cite{Beacom:2002cb} and from searches for the appearance of $\bar{\nu}_e$ at KamLAND~\cite{Eguchi:2003gg}. In some cases, constraints can also be placed from the observations of SN 1987A. 

The 24 neutrino events associated with SN 1987A marked the first (and thus far only) detection of neutrinos from a supernova. Although the modest number of detected events limit the utility of any extensive statistical analysis, some general features have been confirmed.  Specifically, the number of events detected from SN 1987A agree well with expectations.  From this, we can conclude that $\bar{\nu}_e$'s do not significantly decay, at least over timescales of less than $\sim 1.7\times10^5~$years, which limits the corresponding rest-frame lifetime to $\tau_{\bar{\nu}_e} \gsim 0.7 \, {\rm days} \times \left(\frac{m_{\bar{\nu}_e}}{0.1\,{\rm eV}}\right)\left(\frac{10 \, {\rm MeV}}{E_{\bar{\nu}_e}}\right)$.

The constraints from SN 1987A are most relevant for the case of an inverted mass hierarchy, in which $\nu_3$ is the lightest mass eigenstate and, due to the small value of $\theta_{13}$, has a very small projection onto $\nu_e$. Even if $\theta_{13}$ is taken to have the largest experimentally allowed value, the decay scenario where both $\nu_1\rightarrow\nu_3+X$ and $\nu_2\rightarrow\nu_3+X$ leads to about a factor of 10 suppression in the number of expected events.  As this suppression was not observed in SN 1987A, we can conclude that the timescale for this process (for $E_{\nu} \sim 10$ MeV) must also be at least $\sim 1.7\times10^5~$years.  The baseline to the Large Magellanic Cloud (where 1987A occurred) is extremely long at approximately 50~kpc.  As this study is focused on galactic supernova with a baseline of roughly 10~kpc, this scenario of neutrino decay can be excluded from our analysis. Within the context of the normal hierarchy, however, SN 1987A does not strongly constrain the possibility of decays of one neutrino mass eigenstate to another.


In the decay of a neutrino, helicity can either be conserved or flipped.  If the masses are quasi-degenerate ($m_i-m_j \ll m_i, m_j$), then the conserved helicity decay is strongly favored. In this case, the daughter neutrino receives nearly all of the parent's energy, and the overall spectrum is left nearly unchanged. If the masses are instead hierarchical ($m_i-m_j \sim m_i$), then a larger fraction of the parent energy can be lost to the $X$ particle, leading to a degraded spectrum of the daughter species relative to that of the parent. 

The details of the resulting spectrum further depend on whether the neutrinos are Majorana or Dirac particles. If the neutrinos are Majorana particles, then the daughters with flipped helicity will be detectable as a corresponding anti-neutrino.  With the parent (daughter) energy labeled as $E_p\,(E_d)$, the daughter spectrum will be scaled down by a factor of $(E_p-E_d)/E_p^2$ relative to that of the parent~\cite{Ando:2004qe}.  The daughters with conserved helicity will also have a spectrum found by scaling that of the parent, but this time by a factor of $E_d/E_p^2$~\cite{Ando:2004qe}. If the neutrinos are Dirac particles, then the daughter neutrinos with flipped helicity will be sterile and thus undetectable. The daughters with conserved helicity will again have a spectrum produced by scaling the parent spectrum by a factor of $E_d/E_p^2$. Thus in such a decay scenario, not only will the energy spectrum be degraded, but also the total number of neutrinos/anti-neutrinos will be suppressed. 

\subsection{Quantum Decoherence}

Within the context of standard quantum mechanics, a pure state will indefinitely remain in that state, never oscillating into a superposition or mixture of states. It has been suggested, however, that quantum fluctuations of the gravitational field could potentially alter this conclusion. Short lived microscopic black holes, for example, could appear as part of the space time foam, leading to the loss of quantum information as particles move across their event horizons, and transforming pure states into a mixed ones~\cite{qd}. 

Among neutrinos, one might hope to observe quantum decoherence through the loss of flavor information taking place over very long baselines~\cite{qdneu,longbaseline2}. The prediction for this phenomenon is the evolution of neutrino flavors toward a ratio of $\nu_e:\nu_{\mu}:\nu_{\tau} = 1/3:1/3:1/3$, for both neutrinos and anti-neutrinos, regardless of their initial conditions. If this evolution takes place over kiloparsec distance scales, it may be possible to observe with the next galactic supernova.

\section{Neutrinos From Supernovae}
\label{sec:initialFlux}

Over a period of approximately 10 seconds, a supernova produces approximately as many neutrinos as the Sun will generate over its entire life. Approximately 99\% of the gravitational energy associated with a stellar core-collapse supernova is released in the form neutrinos. 

Despite considerable effort spanning several decades in the modeling of core-collapse supernovae, there are still many aspects of the explosion that are not well understood.  In 1998, the Livermoore group produced the most recent simulation that obtained an explosion and followed the evolution through 18~s \cite{Totani:1997vj}.  The model predicts strongly hierarchical time-integrated, average energies: $\left<E_{\nu_e}\right>\,$$\approx\,$12~MeV, $\left<E_{\bar{\nu}_e}\right>\,$$\approx\,$15~MeV, and $\left<E_{\nu_x}\right>\,$$\approx\,$24~MeV, where $\nu_x$ denotes $\nu_{\mu}$, $\nu_{\tau}$, $\bar{\nu}_{\mu}$, and $\bar{\nu}_{\tau}$. This strong hierarchy is useful for distinguishing between different oscillation and decay scenarios, as there are several important neutrino interactions involving $\nu_e$ and $\bar{\nu}_e$ with relatively high energy thresholds that would not produce many events without significant mixing from the harder $\nu_x$ spectra.  The Livermoore group also found luminosity and total energy equipartition between the flavors throughout the evolution.  

\begin{figure}[t]
	\centering
\includegraphics[width=0.775\textwidth]{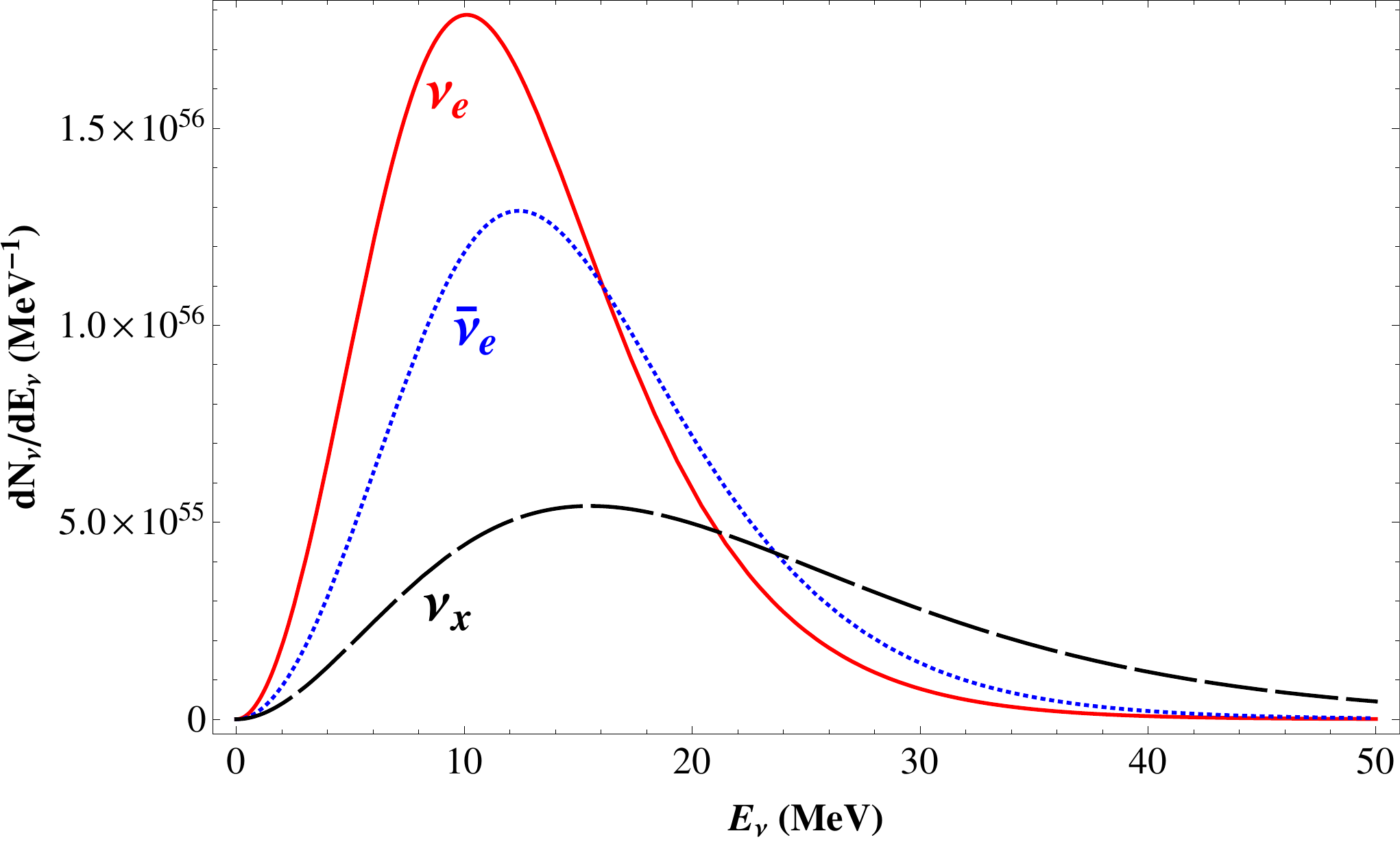} \\
	\caption{The spectrum of neutrinos from a supernova, prior to oscillations, as described by Eq.~\ref{eq:FDspectrum}.}
\label{fig:initial}%
\end{figure}

The time integrated neutrino spectrum is often written as a ``pinched'' Fermi-Dirac distribution:

\begin{equation}
\frac{dN_\nu}{dE_\nu}=\frac{N\,E_\nu^2}{T^4\left(e^{\frac{E_\nu}{T}-\eta}+1\right)},
\label{eq:FDspectrum}
\end{equation}
with $E_\nu$ the energy of the neutrinos, $N$ an overall normalization constant, $T$ the neutrino temperature, and $\eta$ is a chemical potential for the neutrinos. For this spectrum, $\eta$ is a measure of how pinched the low energy and high energy tails would be when compared to a Fermi-Dirac spectrum with the same average energy.  The parameters that we use in our model are chosen to match well with the spectra of the Livermore model, with $T_{\nu_e}=3.5~$MeV, $T_{\bar{\nu}_e}=4~$MeV, $T_{\nu_x}=7~$MeV, $\eta_{\nu_e}=2.08$, $\eta_{\bar{\nu}_e}=2.5$, and $\eta_{\nu_x}=0$~\cite{Totani:1997vj}. These spectra are shown in Fig.~\ref{fig:initial}.

Although the Livermoore group did obtain an explosion with their simulation, the model neglected neutrino processes that are now thought to be important.  More recent simulations which account for these processes fail to produce an explosion and only follow the supernova evolution for a maximum of $\sim$1~s (see Table III of Ref.~\cite{Horiuchi:2008jz} for references to many of these  models).  As discussed in Ref.~\cite{Horiuchi:2008jz}, luminosity equipartition between $\nu_e$ and $\bar{\nu}_e$ seems to hold for most models, but $\nu_x$ luminosity can vary by a factor of 2 in either direction depending on the model and evolutionary phase of the supernova.  Additionally, most of these models produce a somewhat less pronounced energy hierarchy than that found in the Livermoore model.  For example, the Garching Group reports $\left<E_{\nu_e}\right>\approx12$~MeV, $\left<E_{\bar{\nu}_e}\right>\approx 15$~MeV, and $\left<E_{\nu_x}\right>\approx 18$~MeV \cite{Raffelt:2003en}.

\section{The Detection of Supernova Neutrinos}
\label{detect}      

In this section, we summarize the primary detection techniques used to potentially detect the neutrinos from a nearby supernova.

\subsection{Water Cerenkov Detectors}
When charged particles such as electrons travel faster than the local speed of light, they emit radiation which can be detected. Currently, the largest water Cerenkov detector is Super Kamiokande. The signal from an individual supernova is a short lived event ($\sim$10~s), which permits higher background rates, allowing Super Kamiokande to use its full fiducial mass of 32~kilo-tons (kton), significantly larger than any other existing neutrino detector, although detectors at the 500~kton scale have been proposed (UNO \cite{Tonazzo:2007zza}, MEMPHYS \cite{Tonazzo:2007zza}, and Hyper-Kamiokande \cite{HyperK}). The interactions relevant for the observation of supernova neutrinos by water Cerenkov detectors include:

\subsubsection{Inverse Beta Decay: $\bar{\nu}_e+p\rightarrow n+e^+$}

For any target material with a large number of free protons (those not bound in a nucleus), inverse beta decay events will typically dominate the signal.  This is due in large part to the relatively large cross section and the low energy threshold ($\sim$1.8~MeV).  This reaction has been studied extensively both experimentally and theoretically.  We use the cross section from Ref. \cite{Strumia:2003zx} which includes radiative corrections and an exact treatment of the kinematics. 

A proposal has been made to dissolve approximately 0.2\% gadolinium within the water of Super Kamiokande with the purpose of tagging the neutrons produced by inverse beta decay \cite{Beacom:2003nk}. These neutrons quickly thermalize and are captured on the gadolinium nucleus with a 90\% efficiency, producing a cascade of detectable photons \cite{Kibayashi:2009ih}.



\subsubsection{Electron Elastic Scattering: $\nu +e^-\rightarrow \nu+e^-$}


Although all flavors of neutrinos can undergo the process of elastic scattering with electrons, the cross sections vary for the different flavors.  This can be traced to the fact that for electron neutrinos both charged current and neutral current processes contribute to the cross sections.  Supernova neutrinos have energies far below the mass of the muon, so only the neutral current process contributes to the cross section for $\nu_x$.  The approximate ratios of the total cross sections are $\sigma_{\nu_e}:\sigma_{\bar{\nu}_e}:\sigma_{\nu_x}:\sigma_{\bar{\nu}_x}\simeq 1:0.42:0.16:0.14$ \cite{GiuntiBook2007}.  

Although there are copious numbers of electrons in any target material, the cross section is typically much smaller than for interactions with protons or neutrons, causing the total event rate to be dominated by nuclear scattering. Electron scattering events are strongly forward peaked (due to the small electron mass), however, providing a means by which to separate these events from the large rate of inverse beta decay. If the two signals can indeed be separated then this process could provide an opportunity to measure the $\nu_e$ component of the neutrino spectrum. 

\subsubsection{Scattering With Oxygen Nuclei: $\bar{\nu}_e+^{16}${\rm O}~$\rightarrow\,^{16}${\rm N}$+e^+$,~~$\nu_e+^{16}${\rm O}~$\rightarrow\,^{16}${\rm F}$+e^-$,~~$\nu+${\rm O}~$\rightarrow\nu+${\rm O}$^*$}
Supernova neutrinos can also interact with oxygen nuclei in water Cerenkov detectors.  Both $\nu_e$ and $\bar{\nu}_e$ undergo charge current interactions with oxygen nuclei and have thresholds of 15.4~MeV and 11.4~MeV, respectively.  The signals from the resulting positrons and electrons are indistinguishable from each other and from inverse beta decay positrons in Super Kamiokande, so these reactions can not be uniquely identified.  Additionally, the high thresholds mean that unless the harder $\nu_x$ spectra mixes significantly with the $\nu_e$ and $\bar{\nu}_e$ (through oscillations or other exotic physics), the oxygen charged current signal will be quite low.


All neutrino flavors can undergo neutral current reactions with oxygen nuclei, which can yield detectable $\beta$ and $\gamma$ particles in the subsequent decays of the excited nucleus. The most interesting of these decay chains involve excited states of $^{15}$O and $^{15}$N .  These states are only able to decay through the emission of a $\gamma$ with an energy in the range of 5~MeV to 10~MeV and are detectable as peaks above the smooth background (see Figure 2 of Ref.~\cite{oxygenpeaks}).  This oxygen reaction was suggested as a supernova neutrino signal in Ref.~\cite{oxygensn}, where they estimated 300 oxygen events over a smooth background of roughly 270 inverse beta decay events in this energy window.

\subsection{Scintillating Detectors}
Scintillating detectors generally contain an organic oil that is composed primarily of carbon and hydrogen.  Currently, KamLAND  and LVD are the largest scintillating detectors with a mass on the order of 1~kton~\cite{Bandyopadhyay:2003ts,Agafonova:2006fz}.  Although these detectors are much smaller than Super Kamiokande, scintillators typically have much lower energy thresholds and higher light output allowing them to compete with Super Kamiokande in certain arenas.  Proposals for larger scintillators include LENA (50~kton) \cite{Marrodan Undagoitia:2006re} and  Hanohano (10 kton) \cite{Learned:2007zz}.

As scintillators contain many free protons, inverse beta decay provides a strong supernova signal in these detectors.  These events can be tagged in scintillating detectors as the positron produced annihilates with an electron producing a pair of photons that are above the energy detection threshold.  Additionally, there will be a delayed coincidence signal from a 2.2~MeV photon produced when the neutron captures onto a proton producing deuterium.  As with water Cerenkov detectors, electron elastic scattering provides a much smaller signal.  We will now briefly describe the other important interactions for scintillators.

\subsubsection{Scattering With Carbon Nuclei: $\bar{\nu}_e+^{12}${\rm C}~$\rightarrow\,^{12}${\rm B}$+e^+$,~~$\nu_e+^{12}${\rm C}~$\rightarrow\,^{12}${\rm N}$+e^-$,~~$\nu+${\rm C}~$\rightarrow\nu+${\rm C}$^*$}
At supernova neutrino energies, both $\nu_e$ and $\bar{\nu}_e$ can undergo charged current reactions.  The $\nu_e$ reaction has a threshold of 17.3~MeV and produces $^{12}$N which subsequently decays by emitting a $e^+$ with a mean lifetime of 11.0~ms \cite{Bandyopadhyay:2003ts}.  The emitted positron annihilates producing a gamma ray pair that give a delayed coincidence signal for tagging these events.  Similarly, the $\bar{\nu}_e$ reaction has threshold of 14.4~MeV and produces two photons from the emitted positron annihilation.  Also, the $^{12}$B produced in the reaction decays by emitting an electron with a mean lifetime of 20.2~ms, providing an additional coincidence signal \cite{Bandyopadhyay:2003ts}.  With such high thresholds, these reactions depend critically on any mixing from the higher energy $\nu_x$ spectra. 


Neutrinos with energies above the 15.11~MeV threshold can excite $^{12}$C into a state that decays through emission of a distinctive 15.11~MeV, mono-energetic photon.  This signal will convey no spectral information about the neutrinos since all interactions produce the same mono-energetic photon.  This process is neutral current so all flavors contribute equally to the cross section.  Once again, the high threshold limits the interactions to mainly higher energy neutrinos. We have taken the cross section for these reactions from Ref.~\cite{Fukugita:1988hg}.

\subsubsection{Proton Elastic Scattering: $\nu+p\rightarrow\nu+p$}


The cross section for this process is fairly large and all flavors contribute almost equally.  Despite the fact that the scattered protons will undergo a significant amount of quenching due to ionization, this process can still produce a large signal for a supernova \cite{Beacom:2002hs}.  Although the proton elastic scattering would provide additional statistics, we will only use the 15.11~MeV carbon peak for our neutral current signal.  This is because extracting the peak from the smooth background should be quite easy, while separating the proton elastic scattering from the other signals would prove more challenging.

\subsection{Liquid Argon Detectors}

ICARUS is currently the largest liquid argon detector with a mass of 600~tons, and is planned to be scaled up to 3~kton~\cite{Badertscher:2004ij}. In ICARUS, ionization electrons from liquid argon are detected as they travel along a uniform electric field, providing accurate 3-dimensional event tracking and energy reconstruction \cite{:2008sz}.   One of the most attractive features of liquid argon detectors is a strong sensitivity to $\nu_e$ because of the relatively large cross section and low energy threshold.  An additional benefit is that the charged current and neutral current reactions can all be distinguished by gamma emission from the excited nuclei.  With electron elastic scattering events producing no coincident photons, all four of the primary reactions can be separated, in principle.  Two proposed detectors at the 100~kton scale are LANNDD \cite{Cline:2001pt} and GLACIER \cite{Rubbia:2009md}. 

\subsubsection{Scattering With Argon Nuclei: $\nu_e+^{40}${\rm Ar}~$\rightarrow ^{40}${\rm K}$^{\star}+e^-$,~~$\bar{\nu}_e+^{40}${\rm Ar}~$\rightarrow ^{40}${\rm Cl}$^{\star}+e^+$,~~$\nu+${\rm Ar}~$\rightarrow ${\rm Ar}$^{\star}$}

Supernova $\nu_e$ and $\bar{\nu}_e$ both produce charged current reactions in argon with thresholds of 2.8~MeV and 7.9~MeV, respectively \cite{SajjadAthar:2004yf}.  The distinctive photon peaks produced by the de-excitation of the resulting $^{40}${\rm K}$^{\star}$ or $^{40}${\rm Cl}$^{\star}$ nucleus enables these events to be uniquely identified.  The first nuclear excited state of argon is quite low, at 1.46~MeV \cite{GilBotella:2003sz}, making neutral current reactions a particularly strong signal for supernova neutrinos.  The subsequent decay of the argon nucleus will again provide photon signatures to tag these events, but without a corresponding electron or positron.  The cross sections for charged current events can be found in Ref.~\cite{SajjadAthar:2004yf} and neutral current reactions in Ref.~\cite{GilBotella:2003sz}.

\subsubsection{Electron Elastic Scattering: $\nu +e^-\rightarrow \nu+e^-$}

As with previously discussed detector types, electron elastic scattering can also be observed in liquid argon detectors such as ICARUS.

\section{Event Rates and Results}
\label{results}

Supernova neutrinos emerge from their neutrinospheres deep in the core of the progenitor star and must travel from the extremely dense core to the outer surface of the star.  As they travel along this density gradient, they undergo oscillations that are dominated by two resonance regions: a lower density region (L-resonance) that depends on $\Delta m_{12}^2$ and $\theta_{12}$, and a higher density region (H-resonance) that depends on $\Delta m_{13}^2$ and $\theta_{13}$.  The L-resonance occurs in the neutrino sector independently of the mass hierarchy, whereas the H-resonance is significant for neutrinos only if the hierarchy is normal and for anti-neutrinos if the hierarchy is inverted.

The adiabacity of these resonances determines the amount of mixing between the two mass eigenstates. In general, this depends on the matter density gradient, the mixing angle, the mass difference, and energy of the neutrino.  At supernova energies, the L-resonance is perfectly adiabatic, leading to essentially full conversion.  The H-resonance is also perfectly adiabatic if $\sin ^22 \theta_{13}> 10^{-3}$ (referred to as $\theta_{13}$ large), but will be perfectly non-adiabatic ({\it ie.} absent) if $\sin ^22 \theta_{13}< 10^{-5}$ (referred to as $\theta_{13}$ small).  For intermediate values of $\theta_{13}$, the conversion probability at the H-resonance is more complicated and depends on energy.  For simplicity, we will only examine the two boundary cases of perfectly adiabatic and perfectly non-adiabatic for the H-resonance.  For a detailed discussion of oscillations and flavor conversions in supernova neutrinos, see Ref.~\cite{Dighe:1999bi}.

Once the neutrinos emerge from the remnant star, if there is no intervening exotic physics, the mass eigenstates would propagate unchanged to be detected at the earth.  We will present a method for distinguishing this case from a couple of exotic physics scenarios.  




The number of events from a given reaction can be written as 
\begin{equation}
N=N_t\int^{E_{max}}_{E_{min}}\Phi(E_\nu)\sigma(E_\nu)dE_\nu
\label{eq:numberOfEvents}
\end{equation}
with $N_t$ the number of targets in the detector, $\Phi(E_\nu)$ is the flux of the relevant neutrino at the detector, and $\sigma(E_\nu)$ is the cross section for the process.  The limits of the integration are set by the relevant range of neutrino energies.  We have also assumed 100\% detection efficiency and neglected the effects of energy resolution.

\subsection{Results For Water Cerenkov Detectors (Super Kamiokande)}

In Table~\ref{tab:SuperKNormEvents}, we present the numbers of events predicted to be observed from a 10 kpc distant supernova in the Super Kamiokande experiment. Here we have assumed a normal mass hierarchy and considered four possible exotic physics scenarios. First, we consider the case of decays of $\nu_3$ and $\nu_2$ to $\nu_1$ without any significant energy losses, denoted by ``Decay (Full Ener.)''. The columns labeled ``Decay (Majorana)'' and ``Decay (Dirac)'' represent the decay scenarios in which the neutrino masses are hierarchical, leading to a degraded energy spectrum for the daughter neutrinos. If the neutrino masses are quasi-degenerate then the daughter spectrum is unaffected (full energy), independent of whether neutrinos are Majorana or Dirac particles.  Results are shown for both small and large values of $\theta_{13}$. In each decay scenario, we assume that the propagation time significantly exceeds the (boosted) lifetimes of the decays. Finally, the lowest column of the table denotes the case in which all flavor information is lost due to quantum decoherence.



\begin{figure}[!t]
	\centering
\includegraphics[width=0.74\textwidth]{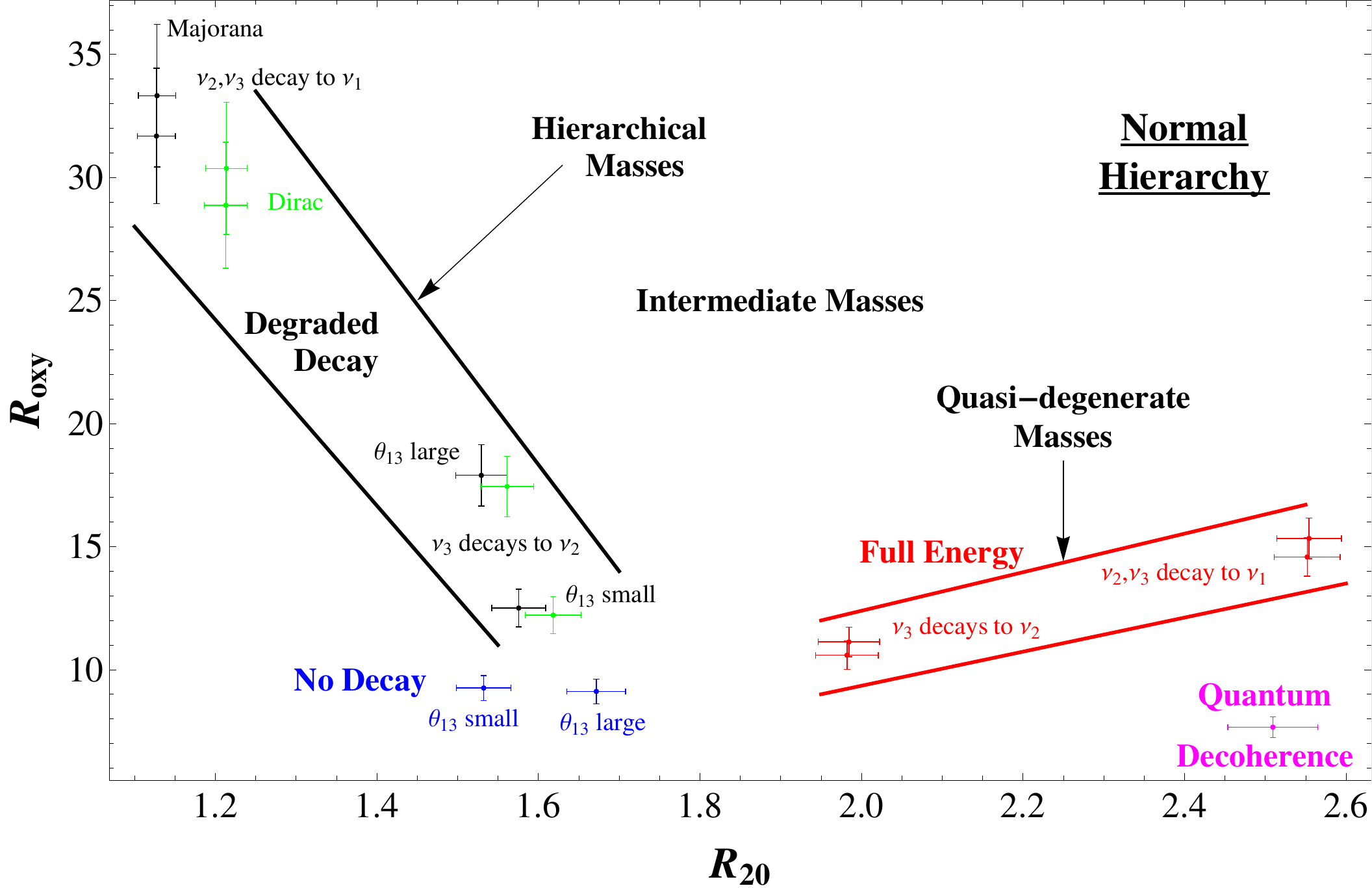} \\
\includegraphics[width=0.74\textwidth]{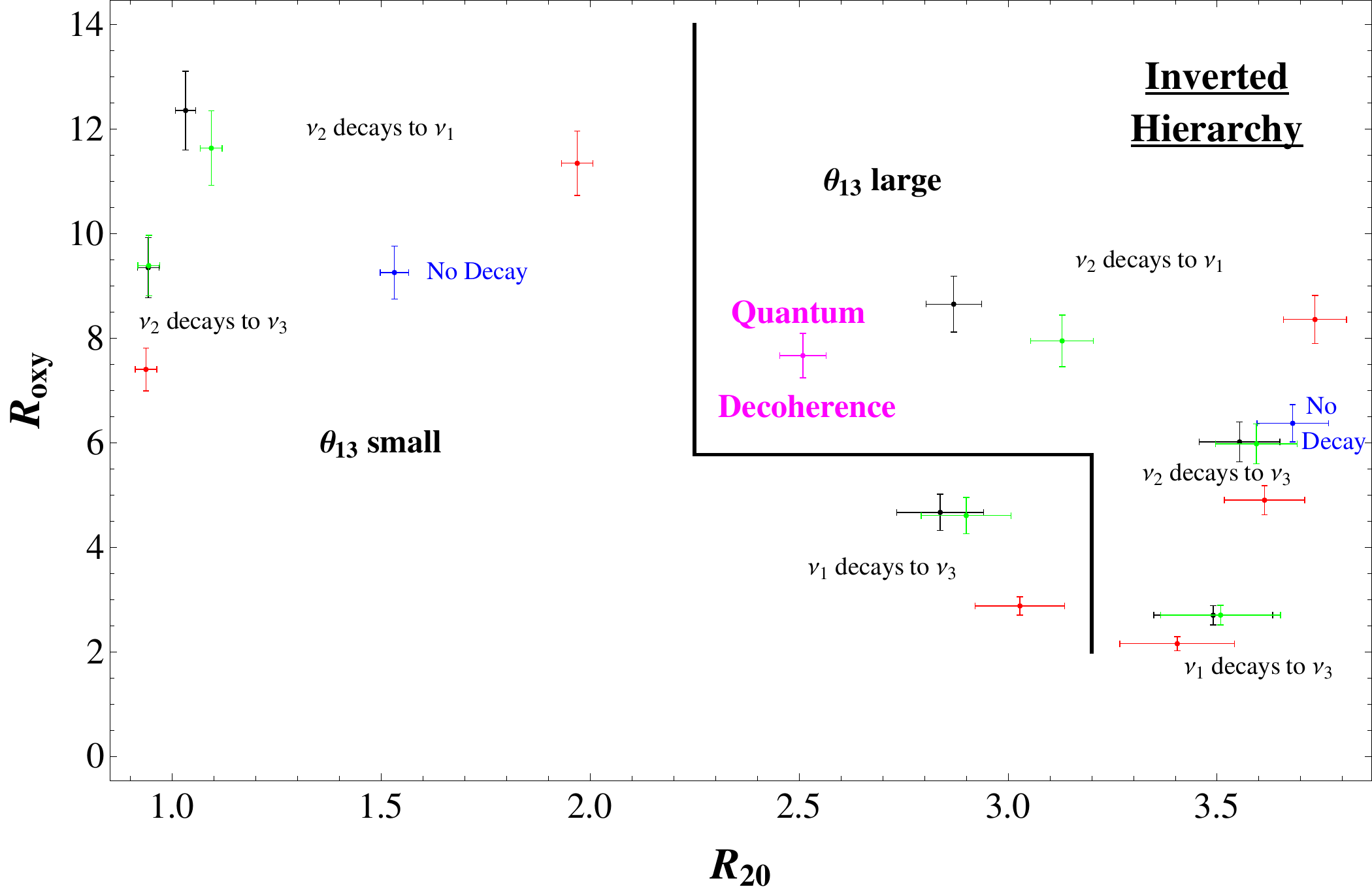} \\
\caption{Event class ratios at Super Kamiokande for the normal (top) and inverted (bottom) mass hierarchies.  $R_{{\rm oxy}}$ is the ratio of total continuum events (including inverse beta decay, electron elastic scattering, and oxygen charged current events) below 20~MeV to oxygen peak events.  $R_{20}$ is the ratio of total continuum events above 20~MeV to the total continuum events below 20~MeV.  The error bars represent the $1\sigma$ statistical uncertainties. Throughout this paper, black and green error bars denote Majorana and Dirac decays, respectively, while red error bars denote full energy decays and blue indicate no decay.}
\label{fig:SuperKPlotNorm}%
\end{figure}

In the upper frame of Fig.~\ref{fig:SuperKPlotNorm}, we present these results in terms of ratios of event types. The first of these, $R_{20}$, is the ratio of total continuum events (including inverse beta decay, electron elastic scattering, and oxygen charged current) above 20~MeV to the total continuum events below 20~MeV.  The second, $R_{{\rm oxy}}$, is the ratio of total continuum events below 20~MeV to oxygen peak events.

\begin{table}[!t]
	\centering
		\begin{tabular}{|l|c|c|c|c|c|c|c|c|c|c|c|c|c|c|c|}
		\cline{2-15}
		\multicolumn{1}{c|}{}		& \multicolumn{4}{c}{$\bar{\nu}_e+p\rightarrow n+e^+$} & \multicolumn{4}{|c}{$\nu+e^{-}\rightarrow \nu+e^{-}$}& \multicolumn{4}{|c|}{$O$ charged current}& \multicolumn{2}{|c|}{$O$ peaks}\\
				\cline{2-15}
		 \multicolumn{1}{c|}{} & \multicolumn{2}{c|}{\scriptsize $E<20$~MeV} & \multicolumn{2}{c|}{\scriptsize $E>20$~MeV} & \multicolumn{2}{c|}{\scriptsize $E<20$~MeV} & \multicolumn{2}{c|}{\scriptsize $E>20$~MeV}& \multicolumn{2}{c|}{\scriptsize $E<20$~MeV} & \multicolumn{2}{c|}{\scriptsize $E>20$~MeV} & \multicolumn{2}{c|}{}\\
		\multicolumn{1}{c}{} & \multicolumn{2}{|c|}{$\theta_{13}$} & \multicolumn{2}{|c|}{$\theta_{13}$} & \multicolumn{2}{|c|}{$\theta_{13}$} & \multicolumn{2}{|c|}{ $\theta_{13}$} & \multicolumn{2}{|c|}{ $\theta_{13}$} & \multicolumn{2}{|c|}{ $\theta_{13}$}& \multicolumn{2}{|c|}{ $\theta_{13}$}\\
		\cline{2-15}
		\multicolumn{1}{c|}{Scenario} & {\footnotesize  small}& {\footnotesize large} & {\footnotesize small} &{\footnotesize large}& {\footnotesize small}& {\footnotesize large} & {\footnotesize small} &{\footnotesize large} & {\footnotesize small}& {\footnotesize large} & {\footnotesize small} &{\footnotesize large}& {\footnotesize small} &{\footnotesize large}\\
		\cline{1-15}
Standard &  3121 & 3063 & 4460 & 4627 & 239 & 238 & 72 & 83 & 67 & 76 & 721 & 935 & 370 & 370\\
Decay (Full Ener.) & 5154 & 4896 & 12573 & 11945 & 379 & 365 & 113 & 109 & 147 & 139 & 1821 & 1730 & 370 & 370\\
Decay (Majorana) & 4246 & 4034 & 4344 & 4127 & 227 & 219 & 61 & 58 & 78 & 74 & 728 & 692 & 137 & 137\\
Decay (Dirac) & 3729 & 3543 & 4093 & 3889 & 197 & 191 & 57 & 55 & 73 & 69 & 704 & 669 & 132 & 132 \\
Quant.~Decoh. & 2525 & 2525 & 6161 & 6161 & 242 & 242 & 72 & 72 & 72 & 72 & 892 & 892 & 370 & 370 \\ 	
		 	\cline{1-15}
		\end{tabular}
	\caption{The number of events in the normal hierarchy detected in Super Kamiokande for a supernova at 10~kpc.  The decay scenario is $\nu_2,\nu_3\rightarrow\nu_1$.  Here $E$ refers to the kinetic energy of the electron (which is the detectable signal) rather than the energy of the neutrino.  The Majorana and Dirac rows show the number of events for neutrinos of that type if the neutrino masses are hierarchical, producing a degraded spectrum for the daughter neutrinos. If the neutrino masses are quasi-degenerate, then the daughter spectrum is unaffected (full energy), independent of whether neutrinos are Majorana or Dirac particles.   The small refers to $\sin ^22 \theta_{13}< 10^{-5}$ and large refers to $\sin^2 2\theta_{13} > 10^{-3}$. See the text for additional initial neutrino parameters. }
	\label{tab:SuperKNormEvents}
\end{table}

The general features of the plot can be understood as follows. As inverse beta decay events dominate the continuum signal, the behavior of $R_{20}$ is mainly determined by the $\bar{\nu}_e$ spectrum.  A larger value of $R_{20}$ indicates that there has been a larger transformation of the initial higher energy $\nu_x$ flux to $\nu_{e}$ through either oscillations in the progenitor star or through exotic physics.  The oxygen peaks, in contrast, are produced through neutral current reactions of the high energy neutrinos.  The number of neutral current events will be the same as long the number and spectrum of the daughter neutrinos are the same as for the parent neutrinos.  For this reason, the largest affect on $R_{{\rm oxy}}$ comes from scenarios where the decay spectra are degraded.  In this case, the daughter neutrino only receives a fraction of the parent neutrino's energy, so fewer high energy neutrinos exist after the decay.  This leads to a large increase in $R_{{\rm oxy}}$ since the number of oxygen peak events decreases significantly.  

At this time, several points pertaining to Fig.~\ref{fig:SuperKPlotNorm} should be noted.  The quantum decoherence signal is well separated from the other scenarios in either hierarchy.  In the inverted hierarchy we have shown the three decay scenarios that are consistent with the supernova 1987A. For the normal hierarchy, only two decay scenarios are presented: $\nu_2,\nu_3\rightarrow\nu_1$ and $\nu_3\rightarrow\nu_2$.  Although other decay scenarios are possible, those shown predict maximum and minimum values of the plotted ratios.  Thus any (full energy) decay scenario will be bounded by the four red points plotted in the figure, and any degraded decay will fall between those shown on the plot. This means that if the neutrinos do decay, this plot should be able to distinguish between the quasi-degenerate (full energy decay) and hierarchical (degraded energy decay) neutrino mass spectrum.  It will be more difficult to discriminate between Majorana and Dirac cases, however. Such measurements may even be capable of distinguishing between small and large values of $\theta_{13}$ at approximately the $2\sigma$ level.


If the neutrino mass hierarchy is inverted, the situation is somewhat more complicated. In particular, predictions for the ratios shown in Fig.~\ref{fig:SuperKPlotNorm} depend significantly on the value of $\theta_{13}$. Without knowledge of this quantity, most decay and decoherence scenarios we have considered cannot be identified using the ratios shown in Fig.~\ref{fig:SuperKPlotNorm}. The exception to this conclusion are decays of the type, $\nu_1\rightarrow\nu_3$, which can potentially be distinguished.

To overcome this challenge in the case of the inverted hierarchy, we consider signals from neutrinos, rather than anti-neutrinos. Unlike anti-neutrinos, the spectrum of neutrinos emerging from the remnant star is independent of the value of $\theta_{13}$ in the inverted hierarchy. Neutrino detection, however, is not the strength of the water Cerenkov detectors. Electron elastic scattering events, which provide the best measurement of the $\nu_e$ flux, are far less numerous than those from inverse beta decay (see Table~\ref{tab:SuperKNormEvents}). They can, however, be separated from the large rate of inverse beta decay events by introducing an angular cut. 

\begin{figure}[!t]
	\centering
\includegraphics[width=0.74\textwidth]{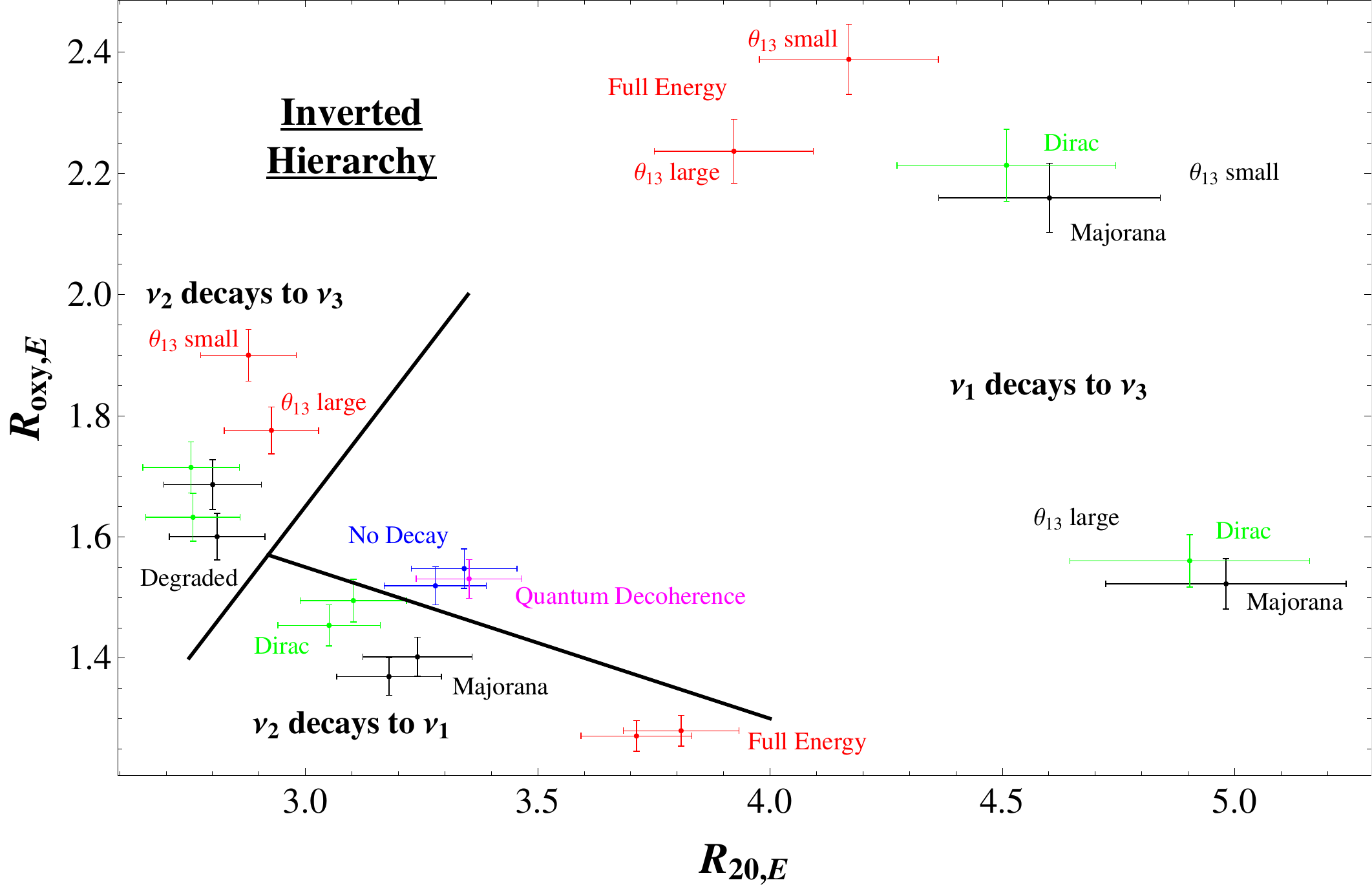} \\
	\caption{Event class ratios at a future large volume (500 kton) water cerenkov detector in the case of the inverted mass hierarchy. $R_{{\rm oxy,E}}$ is the ratio of oxygen peak events to elastic scattering events below 20~MeV.  $R_{20,E}$ is the ratio of elastic scattering events below 20~MeV to the elastic scattering events above 20~MeV. Again, the error bars represent the $1\sigma$ statistical uncertainties.}
\label{fig:SuperKPlotInv2}%
\end{figure}

In Fig.~\ref{fig:SuperKPlotInv2}, we show ratios of events chosen to better separate the various scenarios being considered in the case of the inverted mass hierarchy. These ratios are essentially the inverse of the ratios shown in Fig.~\ref{fig:SuperKPlotNorm}, but using electron elastic scattering events in place of the total continuum events (the reason for inverting the ratio is because the elastic scattering rate is quite small, so inverting the ratio gives a larger possible range).   The error bars that are shown are for a future water Cerenkov detector with 500~kton fiducial mass (rather than the 32~kton of Super Kamiokande).



As can be seen from Fig.~\ref{fig:SuperKPlotInv2}, these ratios do provide separation between the different scenarios with the better statistics from a larger detector.  Unfortunately, the error bars for Super Kamiokande would be approximately four times larger, indicating that Super Kamiokande would likely only be able to separate the $\nu_1\rightarrow\nu_3$ decay scenario if the mass hierarchy is inverted. It is also interesting to note that for these ratios, a coincidence in the initial spectra causes the predictions in the no decay and quantum decoherence scenarios to be almost identical.  The quantum decoherence signal is well separated in both frames of Fig.~\ref{fig:SuperKPlotNorm} though, indicating that those ratios provide the best  opportunity to identify this signal, independent of mass hierarchy.

\subsection{Results For Scintillation Detectors (KamLAND)}

\begin{figure}[t]
	\centering
\includegraphics[width=0.74\textwidth]{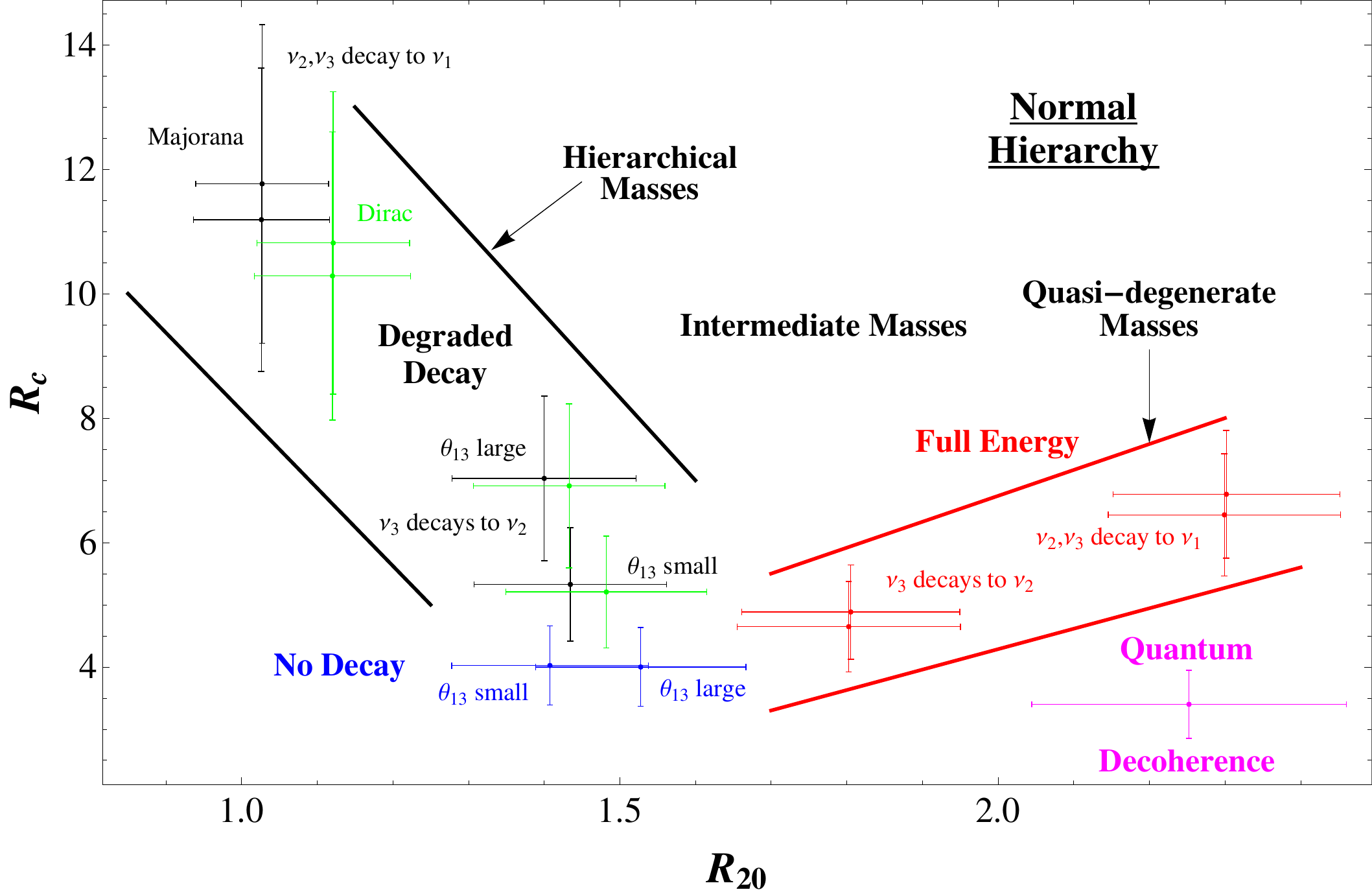} \\
	\caption{Event class ratios in the KamLAND detector in the case of the normal mass hierarchy. $R_{{\rm c}}$ is the ratio of total continuum events (includes inverse beta decay, electron elastic scattering, and carbon charged current) below 20~MeV to the 15.11~MeV gamma ray peak events from the de-excitation of the carbon nucleus.  $R_{20}$ is the ratio of total continuum events above 20~MeV to the total continuum events below 20~MeV. Again, the error bars represent the $1\sigma$ statistical uncertainties.}
\label{fig:KamLANDPlotNorm}%
\end{figure}

Using KamLAND, with a mass of 1~kton, as a representative scintillating detector, we show in Table~\ref{tab:KamLANDNormEvents} the numbers of various types of events, in various exotic physics scenarios. Fig.~\ref{fig:KamLANDPlotNorm}, we again introduce $R_{20}$ as the ratio of continuum events (inverse beta decay, electron elastic scattering, and charged current events on carbon nuclei) above 20~MeV to continuum events below 20~MeV, and compare this to $R_c$, which we define as the ratio of total continuum events below 20~MeV to the number of events in the gamma ray peak at 15.11~MeV from the de-excitation of the carbon nucleus. 

\begin{table}[t]
	\centering
		\begin{tabular}{|l|c|c|c|c|c|c|c|c|c|c|c|c|c|c|c|}
		\cline{2-15}
		\multicolumn{1}{c|}{}		& \multicolumn{4}{c}{$\bar{\nu}_e+p\rightarrow n+e^+$} & \multicolumn{4}{|c}{$\nu+e^{-}\rightarrow \nu+e^{-}$}& \multicolumn{4}{|c|}{$C$ charged current}& \multicolumn{2}{|c|}{$C$ peak}\\
				\cline{2-15}
		 \multicolumn{1}{c|}{} & \multicolumn{2}{c|}{\scriptsize $E<20$~MeV} & \multicolumn{2}{c|}{\scriptsize $E>20$~MeV} & \multicolumn{2}{c|}{\scriptsize $E<20$~MeV} & \multicolumn{2}{c|}{\scriptsize $E>20$~MeV}& \multicolumn{2}{c|}{\scriptsize $E<20$~MeV} & \multicolumn{2}{c|}{\scriptsize $E>20$~MeV} & \multicolumn{2}{c|}{}\\
		\multicolumn{1}{c}{} & \multicolumn{2}{|c|}{$\theta_{13}$} & \multicolumn{2}{|c|}{$\theta_{13}$} & \multicolumn{2}{|c|}{$\theta_{13}$} & \multicolumn{2}{|c|}{ $\theta_{13}$} & \multicolumn{2}{|c|}{ $\theta_{13}$} & \multicolumn{2}{|c|}{ $\theta_{13}$}& \multicolumn{2}{|c|}{ $\theta_{13}$}\\
		\cline{2-15}
		\multicolumn{1}{c|}{Scenario} & {\footnotesize  small}& {\footnotesize large} & {\footnotesize small} &{\footnotesize large}& {\footnotesize small}& {\footnotesize large} & {\footnotesize small} &{\footnotesize large} & {\footnotesize small}& {\footnotesize large} & {\footnotesize small} &{\footnotesize large}& {\footnotesize small} &{\footnotesize large}\\
		\cline{1-15}

Standard &  167 &  164 &  237 &  245 &  17 &  17 &  3.0 &  3.5 &  16 &  19 &  43 &  56 &  50 & 50\\
Decay (Full Ener.) &  276 &  262 &  667 &  634 &  27 &  26 &  4.8 &  4.6 &  35 &  33 &  106 &  101 &  50& 50\\
Decay (Majorana) &  233 &  221 &  230 &  219 &  18 &  18 &  2.6 &  2.5 &  19 &  18 &  44 &  42 &  23 & 23\\
Decay (Dirac) &  201 &  191 &  217 &  206 &  15 &  14 &  2.4 &  2.3 &  17 &  17 &  43 &  41 &  22 & 22\\
Quant.~Decoh. &  135 & 135 & 327 & 327   &  17 & 17  &  3.1 & 3.1  &  17 & 17  &  52 & 52 & 50 & 50	\\
		 	\cline{1-15}
		\end{tabular}
	\caption{The same as Table~\ref{tab:SuperKNormEvents} but for the KamLAND scintillating detector.}
	\label{tab:KamLANDNormEvents}
\end{table}

Again, we find that each of the decay and decoherence scenarios occupies a distinct region in the plot. The error bars here a bit larger than in Fig.~\ref{fig:SuperKPlotNorm} owing to the smaller detector mass of KamLAND compared to Super Kamiokande.  The error bars of Fig.~\ref{fig:KamLANDPlotNorm} would be roughly seven times smaller for the planned scintillating detectors in the 50~kton range, however, giving such future detectors even better resolving power than Super Kamiokande.

As in the case of water cerenkov detectors, if the neutrino mass hierarchy is inverted, oscillations complicate the situation.  In fact, even the next generation scintillating detectors in the 50~kton mass range are not well equipped to distinguish the decay scenarios for the inverted mass hierarchy.  As discussed in the previous section, neutrino oscillations play the largest role in determining the signals in the anti-neutrino sector.  Oscillations play a much more minor role in the neutrino sector providing the hope that the neutrino flux might provide the ability to separate the decay scenarios.

Unfortunately, the charged-current reactions of $\nu_e$ on carbon nuclei has a very high energy threshold of 17.3~MeV, and in the inverted hierarchy, $\nu_2$ will have the softest spectrum.  This means that any scenario in which $\nu_2$ decays will have only minor affects on the ratios.  We also consider the possibility of using electron elastic scattering in the same manner as was done for Super Kamiokande.  This is not practical, however, because electron elastic scattering requires detectors on the order of 500~kton to effectively separate the scenarios (see Fig. \ref{fig:SuperKPlotInv2}), approximately 10 times larger than even the next generation scintillating detectors (which are of similar size to Super Kamiokande). 


The one scenario that would be very easily identified is again $\nu_1\rightarrow\nu_3$.  In this scenario, the charged current reactions on carbon would virtually disappear because nearly all of the high energy neutrinos will be $\nu_3$, which has a very small projection onto $\nu_e$.  


\subsection{Results For Liquid Argon Detectors (ICARUS)}
Both water Cerenkov and scintillating detectors are much more sensitive to $\bar{\nu}_e$, providing good separation ability between decay scenarios in the normal mass hierarchy.  This is because the anti-neutrino sector is best for isolating the decay differences from oscillation effects in the normal mass hierarchy.  The true strength of a liquid argon detector is realized in the inverted mass hierarchy.  In this case, the neutrino sector provides the best opportunity for isolating decay effects. The low energy threshold (2.8~MeV) for charged-current neutrino reactions with Argon nuclei gives ICARUS comparatively much better sensitivity to the $\nu_e$ flux over the other two detector types. Although the final proposed mass for ICARUS is 3~kton~\cite{Badertscher:2004ij}, we use the current detector mass of 0.6~kton as our model case for liquid argon detectors.    

In principle, ICARUS will be able to isolate each of the four types of interactions in the detector.  This is another huge benefit for liquid argon over the other types of detectors.  Some representative results for the number of events in ICARUS for the different scenarios in the inverted mass hierarchy are shown in Table~\ref{tab:IcarusNormEvents}.  The strong neutrino charged-current and neutral current signals are quite noticeable in the table.

\begin{table}[!t]
	\centering
		\begin{tabular}{|l|c|c|c|c|c|c|c|c|c|c|c|}
		\cline{2-11}
		\multicolumn{1}{c|}{}		& \multicolumn{4}{c}{$\nu_e+{}^{40}Ar\rightarrow {}^{40}K^*+e^-$} & \multicolumn{4}{|c}{$\bar{\nu}_e+{}^{40}Ar\rightarrow {}^{40}Cl^*+e^+$}& \multicolumn{2}{|c|}{neutral current}\\
				\cline{2-11}
						 \multicolumn{1}{c|}{} & \multicolumn{2}{c|}{$E<20$~MeV} & \multicolumn{2}{c|}{$E>20$~MeV} & \multicolumn{2}{c|}{$E<20$~MeV} & \multicolumn{2}{c|}{$E>20$~MeV} & \multicolumn{2}{c|}{}\\
		\multicolumn{1}{c}{} & \multicolumn{2}{|c|}{$\theta_{13}$} & \multicolumn{2}{|c|}{ $\theta_{13}$} & \multicolumn{2}{|c|}{$\theta_{13}$} & \multicolumn{2}{|c|}{ $\theta_{13}$}& \multicolumn{2}{|c|}{ $\theta_{13}$}\\
		\cline{2-11}
		\multicolumn{1}{c|}{Scenario} & small& large & small &large & small& large & small &large& small &large\\
		\hline
Standard & 41 & 41 & 68 & 69 & 2.4 & 3.5 & 3.0 & 7.9 & 113 & 113\\
Decay (Full Ener.) & 52 & 52 & 69 & 71 & 3.7 & 4.7 & 6.0 & 10.7 & 113 & 113\\
Decay (Majorana) & 37 & 38 & 68 & 70 & 1.9 & 3.0 & 1.1 & 6.1 & 91 & 91\\
Decay (Dirac) & 34 & 35 & 67 & 68 & 1.9 & 3.0 & 1.1 & 6.1 & 90 & 90\\
Quant.~Decoh. & 40 & 40 & 66 & 66 & 3.0 & 3.0 & 5.7 & 5.7 & 113 & 113\\
		 	\hline
		\end{tabular}
	\caption{The number of events in the inverted hierarchy detected in the 0.6~kton liquid argon detector ICARUS for a supernova at 10~kpc.  The decay scenario is $\nu_2\rightarrow\nu_1$. The labels are similar to Tables~\ref{tab:SuperKNormEvents} and \ref{tab:KamLANDNormEvents}.}
	\label{tab:IcarusNormEvents}
\end{table}

\begin{figure}[!t]
	\centering
\includegraphics[width=0.74\textwidth]{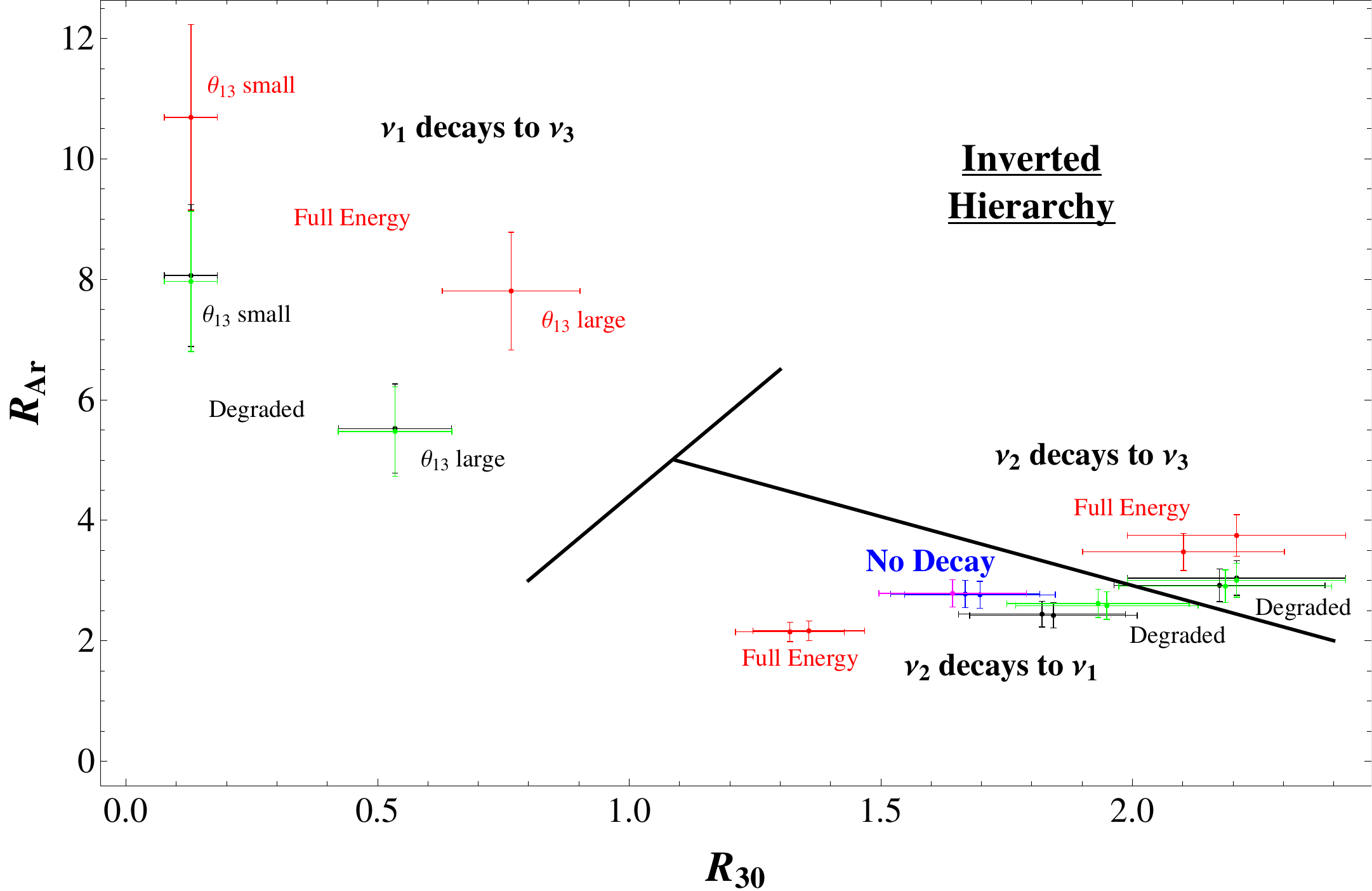} \\
	\caption{Event class ratios in the ICARUS detector in the case of the inverted mass hierarchy. $R_{30}$ is the ratio of charged-current neutrino events above 30~MeV to below 30~MeV, and $R_{{\rm Ar}}$ is the ratio of total neutral current events to the charged-current neutrino events below 30~MeV. The error bars represent the $1\sigma$ statistical uncertainties, for the planned final detector mass of 3~kton (current mass is 0.6~kton). Notice that the quantum decoherence point is coincidentally indistinguishable from the no decay scenario.}
\label{fig:IcarusPlotInv}%
\end{figure}

We again introduce two ratios to help to separate the different decay scenarios.  We define $R_{30}$ as the ratio of charged-current neutrino events above 30~MeV to those below 30~MeV.   We also define $R_{{\rm Ar}}$ as the ratio of total neutral current events to the charged-current neutrino events below 30~MeV.  These ratios are plotted in Fig.~\ref{fig:IcarusPlotInv}.  For clarity, the $1\sigma$ errors bars for the 3~kton mass detector are plotted in the the figure.  In comparing Fig.~\ref{fig:IcarusPlotInv} to the lower frame of Fig.~\ref{fig:SuperKPlotNorm}, we find that, in its final form, ICARUS will already be competitive with the next generation water Cerenkov detectors in the case of the inverted mass hierarchy.  A future liquid argon detector of 100~kton would reduce the error bars by about a factor of 6, providing the best separation power in the inverted hierarchy by a wide margin.

The threshold for anti-neutrino charged current reactions on argon nuclei is not particularly large at 7.5~MeV. However, the much lower threshold of 1.8~MeV for inverse beta decay ensures that both water Cerenkov and scintillating detectors of similar size will have better separation sensitivity in the normal mass hierarchy.  We have calculated the results in the normal mass hierarchy (not shown) for the next generation liquid argon detectors and found similar scenario separation to that of Super Kamiokande (see Fig.~\ref{fig:SuperKPlotNorm}).   

All of the results that we have presented thus far are dependent upon the strong hierarchy of initial average energies within the Livermoore model.  We also calculated similar results to those presented above using the the softer $\nu_x$ spectrum of the Garching model.  Although the statistical uncertainties are increased, we still found that the detectors were able to separate the scenarios.  However, the values of the corresponding ratios between the two models do not match well. If a particular model emerges as the accepted one among the community, then our calculations could very easily be modified to the accepted model.  One way to avoid model dependent conclusions is by focusing on the neutronization burst from the supernova as discussed in the following section.

\section{Neutrinos From The Neutronization Burst}
\label{neutronization}
At the onset of instability, the core begins to collapse, significantly increasing in density.  Once the density of the inner core reaches that of nuclear matter of order ($10^{14}~$g~cm$^{-3}$), the core bounces, producing an outgoing shock wave. The shock wave travels towards the outer layers of the stellar core, losing energy through the photodissociation of nuclei.  The newly created free protons have a large electron capture rate, which leads them to neutronize through the process $e^-+p\rightarrow n + \nu_e$, resulting in a large flux of $\nu_e$.  These neutrinos are initially trapped behind the shock wave, as the density of the core is too high.  Once the shock wave has traveled outwards to the region of the neutrinosphere ($\rho\approx 10^{11}~$g~cm$^{-3}$), however, the neutrinos rapidly escape in what is often termed the neutronization burst.  

The presence and structure of the neutronization peak is one of the most robust features of supernova models.  In Ref.~\cite{Kachelriess:2004ds}, the authors show that several different models using many different progenitor stars produce very similar peaks.  Additionally, the authors examined the dependence of the signal in a megaton water Cerenkov detector on several important astrophysical inputs. They found that differences in the progenitor mass, three possible models for nuclear equation of state for the core, and possible improvements to interaction cross sections all led to changes in event numbers that were small compared to the statistical fluctuations.

The robustness of the neutronization peak across many models provides the opportunity for examining exotic physics scenarios without relying on the details of the underlying supernova model.  One limitation of using the neutronizaton burst peak, however, is that the signal is not particularly bright. The neutronization peak typically has a full width half maximum of 5-7~ms and a peak luminosity of $3.3-3.5\times10^{53}~$erg s$^{-1}$ \cite{Kachelriess:2004ds}.  The total energy released is $\sim$$3\times10^{51}~$erg, compared to the much larger $\sim$$5\times10^{52}~$erg of energy in $\nu_e$ for the entire supernova. As a consequence, very large volume detectors will be required to discriminate between the various exotic physics scenarios being considered here. 

\begin{figure}[!t]
	\centering
\includegraphics[width=0.74\textwidth]{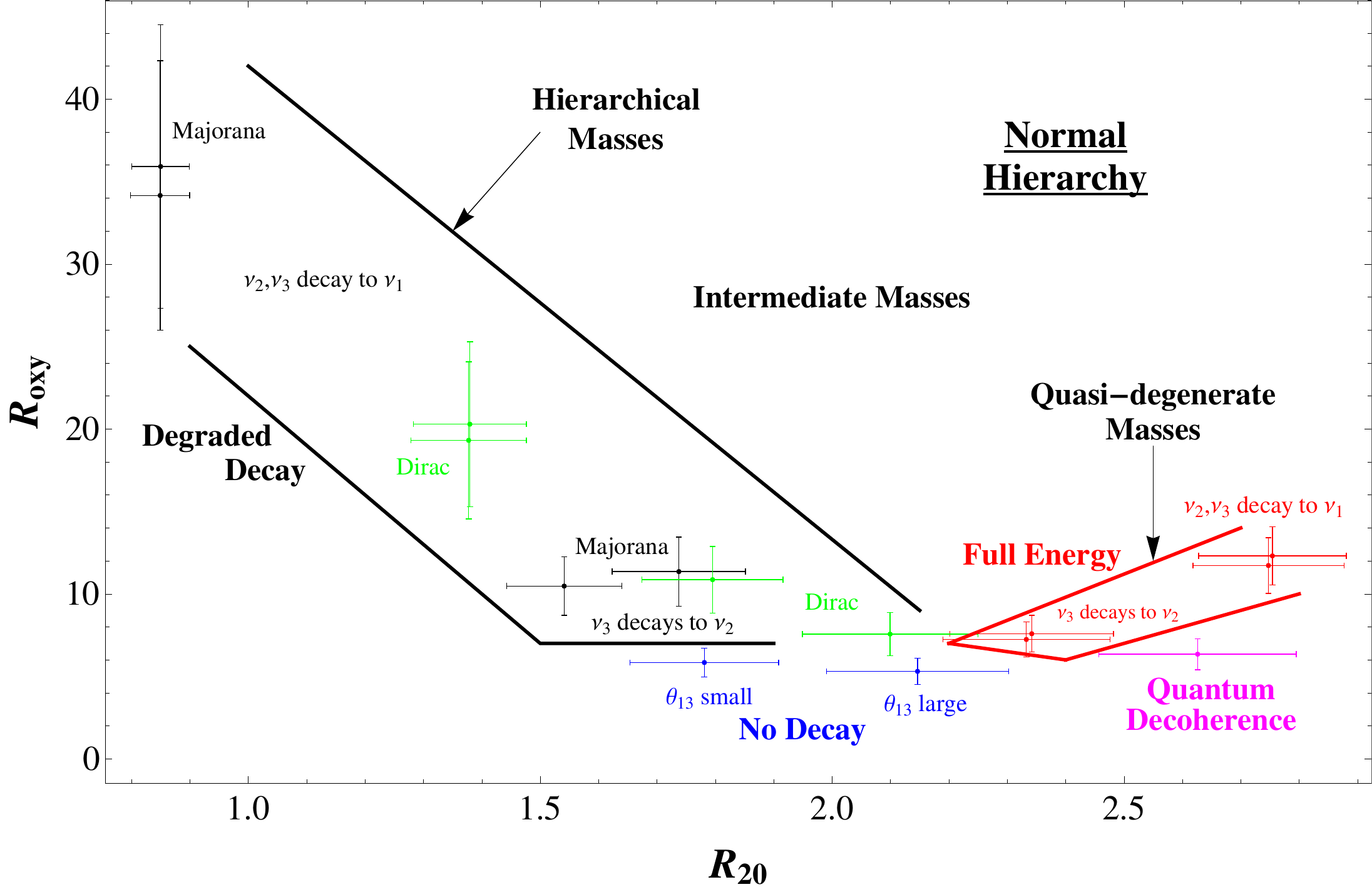} \\
\includegraphics[width=0.74\textwidth]{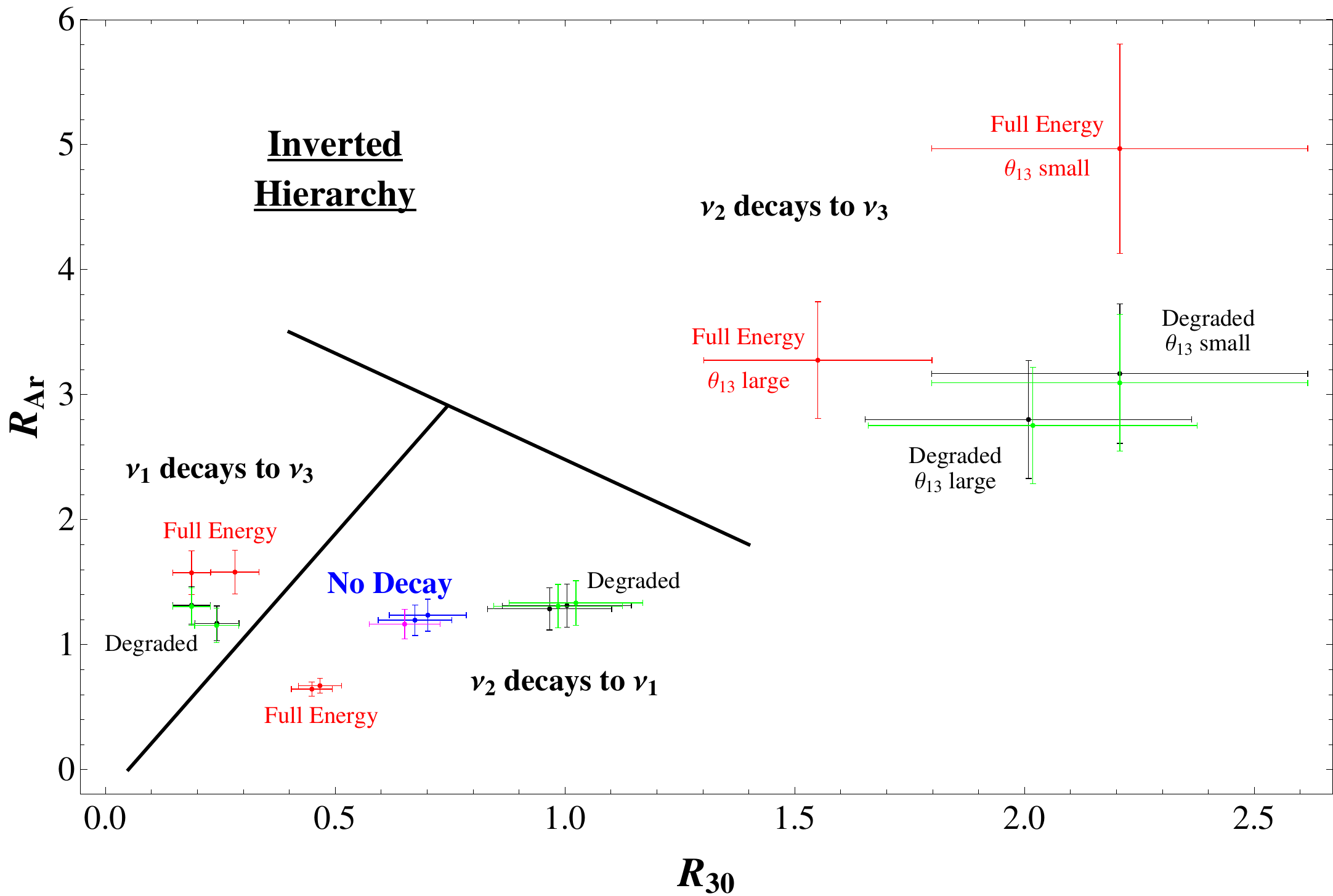} \\
\caption{Event class ratios for the neutronization burst signal.  The ratios are the same as in Fig.~\ref{fig:SuperKPlotNorm} for a 500~kton water Cerenkov detector in the normal hierarchy (top) and as in Fig.~\ref{fig:IcarusPlotInv} for a 100~kton liquid argon detector in the inverted hierarchy (bottom).   The error bars represent the $1\sigma$ statistical uncertainties. }
\label{fig:neutronization}%
\end{figure}

In Ref.~\cite{Ando:2004qe}, the authors examine neutrino decay (for three different decay scenarios) in the case of the normal mass hierarchy with hierarchical masses, in which the neutronization peak can be identified in the inverse beta decay signal in a future water Cerenkov detector.  Similarly, we also find that future planned detectors will possess the ability to clearly separate between our exotic physics scenarios when focusing only on the neutronization burst.  In Fig.~\ref{fig:neutronization}, we plot the same ratios as in the previous sections, but for the neutronization signal only.  The error bars shown are again statistical and correspond to a 500~kton water Cerenkov detector in the normal hierarchy and for a 100~kton liquid argon detector in the inverted hierarchy.  An important feature of this plot is shown in the normal hierarchy with quasi-degenerate masses.  If the both $\nu_2$ and $nu_3$ decay, then the water Cerenkov detector would be able to make a strong distinction between the the Majorana and Dirac nature of the neutrinos.  

\section{Conclusions}
\label{conclusions}
In this article, we have calculated the expected neutrino spectrum and corresponding signals from a galactic core-collapse supernova using the Livermoore model.  We utilize the extremely long baseline to investigate possible signals of various exotic physics models.  By introducing ratios of event types to distinguish between these various scenarios, we eliminate some possible supernova uncertainties (such as total energy released, distance to supernova, etc.).  In the case of a normal mass hierarchy, Super Kamiokande will be able to distinguish between various decay scenarios.  In the case of an inverted mass hierarchy, ICARUS is best able to discriminate between the possible models.  In either mass hierarchy, Super Kamiokande is best equipped to identify the signatures of quantum decoherence.  We also find that future water Cerenkov ($\sim$500~kton) and liquid argon ($\sim$100~kton) detectors will be able to make strong conclusions about these phenomenon in a model independent way by examining the neutrino signal associated with a supernova's neutronization burst.

\bigskip
\bigskip
This work has been supported by the US Department of Energy, including grant DE-FG02-95ER40896, and by NASA grant NAG5-10842.

\end{document}